\documentclass[singlecolumn,epjc3]{svjour3}
\smartqed  
\RequirePackage{graphicx}
\journalname{Eur. Phys. J. A}

\begin{document}

\title{Modeling of charged anisotropic compact stars in general relativity}

\author{
Baiju Dayanandan  \thanksref{e1,addr1}
\and S.K. Maurya \thanksref{e2,addr1}
\and Smitha T. T. \thanksref{e3,addr1}}

\thankstext{e1}{e-mail: baiju@unizwa.edu.om}
\thankstext{e2}{e-mail:sunil@unizwa.edu.om}
\thankstext{e3}{e-mail: smitha@unizwa.edu.om}

\institute{
Department of Mathematical \& Physical Sciences,
College of Arts \& Science, University of Nizwa, Nizwa, Sultanate
of Oman\label{addr1}
}

\date{Received: date / Accepted: date}

\maketitle

\begin{abstract}
A charged compact star model has been determined for anisotropic fluid distribution. We have solved the Einstein's- Maxwell field equations to construct the charged compact star models by using radial pressure, metric function $e^{\lambda}$ and electric charge function. The generic charged anisotropic solution is verified by exploring different physical conditions like, causality condition, mass-radius relation and stability of the solution (via. adiabatic index, TOV equations and Herrera cracking concept). It is observed that the present charged anisotropic compact star is compatible with the star PSR 1937+21. However we also presented the EOS $\rho=f(p)$ for present charged compact star model.
\end{abstract}


\maketitle

\section{Introduction}
The search for theoretical compact star modeling, one of the centerpieces of general relativity, has been a work in progress for over five decades. Over  the  years  the  workers  in general  relativity  had  been using  perfect  fluid as a source  of stellar  structures  to form models for compact stars \cite{1,J1,2,3,4,5,6}. Later on it was observed that for stars with highly dense core do not yield realistic result with the perfect fluid as their source. It was proposed by Ruderman\cite{7} and Canuto\cite{8} that the stars may have high density ranges, when the nuclear matter has anisotropic features. Consequently fluids with the anisotropy in pressure were chosen as source to depict the stellar models in order to improve the results considerably \cite{9,10,11,12,13,14,15,16,17,18,19,20,21,22}. On the other hand gravitational collapse is one of the most dramatic phenomena in universe. When the pressure is not sufficient to balance the gravitational attraction inside the star, the star undergoes sudden gravitational collapse and physical characteristic of star changes dramatically. The presence of Charged compact objects may avert the gravitational attraction is counterbalanced by the repulsive Colombian force in addition to the pressure gradient. There have been several investigations of compact star models in the presence of an electric field in recent years on the Einstein-Maxwell system of equations \cite{23,24,25,26,27,28,29,30}. The presence of charge affects values for redshifts, luminosity and maximum mass for stars astrophysics where the gravitational fields are strong are described by solutions of the coupled Einstein-Maxwell system of equations. More over the charged analogues of the said anisotropic models further improved the situation \cite{31,32,33}. Lake\cite{2} and Herrera et al.\cite{34} have developed the algorithm for all spherical symmetric solutions corresponding to perfect and anisotropic fluid distribution respectively. Later on Maurya et al.\cite{35} has generalized this algorithm for charged anisotropic fluid distributions.  Also, we have given several conditions on physical quantities for constructing the realistic models. Several researchers have been obtained solutions of Einstein's equations in different approaches which can be seen the following recent references\cite{36,37,38,39,40,41,42,43,44}

Recently Piyali et al.\cite{45,46} have obtained the anisotropic solutions by taking radial pressure and mass function but both these solutions are unstable at some points inside the star due to anisotropic factor $\Delta$ is negative. However in the present paper a class of more generalized stellar models are developed in same manner by considering physically suitable expressions of radial pressure, charged intensity and to some extent mass function (by choosing  $e^{\lambda}$ suitably). The models so obtained  are neatly analyzed subject to adequate physical conditions. We found that the present charged anisotropic solution is stable at every point inside the star. In addition to that we have also provided the EOS for the present compact star which has the form $\rho=f(p)$.

The outline of the paper as follows: In section 2, The Einstein-Maxwell equations has been presented including with spherically symmetric metric. The new charged anisotropic solution is derived by help of $e^{\lambda}$, $q$ and $p_r$ in sec. 3. We have mentioned the equation of state (EOS), which is most vital one, in sec. 4. However, In sec. 5 the boundary conditions have been applied to determine the constants. The physical features of the models with stability conditions have been discussed in details in Sec.6. At the last we discuss the result with concluding remarks.

\section{The Einstein-Maxwell field equations for spherical symmetric anisotropic fluid distribution}
Let us consider the spherical symmetric metric in curvature coordinates as :
\begin{equation}
	\label{4}
	ds^2 = - e^{\lambda(r)} dr^2+e^{\nu(r)} dt^2- r^2(d\theta^2 + sin^2\theta d\phi^2) .
\end{equation}

Then the Einstein-Maxwell equations for anisotropic fluid distribution is defined as (by taking G=c=1):

\begin{equation}
	\label{1}
	-8\pi\, (T^i_{j}+E^i_{j})=R^i_{j}-\frac{1}{2}R\,\delta^i_{j}.
\end{equation}

Where the energy momentum tensor ($T^i_{j}$) and electromagnetic field tensor ($E^i_{j}$) can be written as:
\begin{equation}
\label{2}
	T^i_{j} = [(\rho + p_t)u^i\,u_j+(p_r-p_t) \mu^i \mu_j-p_t\,\delta^i_{j}],
\end{equation}

\begin{equation}
\label{3}
	E^i_{j} = \frac{1}{4\pi}[-F^{ik}F_{jk}+\frac{1}{4}\delta\,F^{kl}F_{kl}],
\end{equation}

where $u^i$ is the four-velocity as
$u^i=e^{\nu(r)/2}{\delta^i}_4$ and $\mu^i$ is the unit space like
vector in the direction of radial vector as $\mu^i=e^{\lambda(r)/2}{\delta^i}_1$, $\rho $ is the energy density,
$p_r$ is the pressure in the direction of $\mu^i$ (normal or
radial pressure) and $p_t$ is the pressure orthogonal to
$\mu_i$ (transverse or tangential pressure). Also, anti-symmetric electromagnetic field tensor $F_{ij}$ satisfies the Maxwell equations

\begin{equation}
F_{ik,j} + F_{kj,i} + F_{ji,k} = 0, \label{F12}
\end{equation}

and

\begin{equation}
\frac{\partial}{\partial x^k}({\sqrt -g} F^{ik}) = -4\pi{\sqrt -g}
J^i, \label{max2}
\end{equation}

where, $J^i$ is the four-current vector defined by
\begin{equation}
J^i = \frac{\sigma}{\sqrt g_{44}} \frac{dx^i}{dx^4} = \sigma v^i, \label{J12}
\end{equation}

where $\sigma$ is the charged density.

For static matter distribution the only non-zero component of the
four-current is $J^4$. Because of spherical symmetry, the
four-current component is only a function of radial distance, $r$.
The only non vanishing components of electromagnetic field tensor
are $F^{41}$ and $F^{14}$, related by $F^{41} = - F^{14}$, which
describe the radial component of the electric field. From the Eq.
(\ref{max2}), one obtains the following expression for the
electric field:
\begin{equation}
F^{41} = e^{-(\nu+\lambda)/2} \left[\frac{q(r)}{r^2}\right],
\end{equation}
where $q(r)$ represents the total charge contained within the
sphere of radius $r$, which is defined by
\begin{equation}
q(r) = 4\pi \int_0^r \sigma r^2 e^{\lambda/2} dr = r^2 \sqrt{-F_{14}F^{14}} = r^2 F^{41} e^{(\nu+\lambda)/2}.\label{charge}
\end{equation}
Equation (\ref{charge}) can be treated as the relativistic version
of Gauss's law, which gives

\begin{equation}
\frac{\partial}{\partial r}(r^2 F^{41} e^{(\nu+\lambda)/2}) = - 4\pi r^2 e^{(\nu+\lambda)/2} J^4.
\end{equation}

In view of metric (\ref{4}), The Einstein-Maxwell field Eqs.(\ref{1}),(\ref{F12}), and (\ref{J12}) give the following differential equations as:

\begin{equation}
\label{5}
-8\pi {T^1}_1 = \frac{{\nu}^{\prime}}{r} e^{-\lambda} - \frac{(1 - e^{-\lambda})}{r^2} = 8\pi p_r - \frac{q^2}{r^4},
\end{equation}

\begin{eqnarray}
-8\pi {T^2}_2 = -8\pi {T^3}_3 = \left[\frac{{\nu}^{\prime\prime}}{2} - \frac{{\lambda}^{\prime}{\nu}^{\prime}}{4} + \frac{{{\nu}^{\prime}}^2}{4}
+ \frac{{\nu}^{\prime} - {\lambda}^{\prime}}{2r}\right]e^{-\lambda}  = 8\pi p_t + \frac{q^2}{r^4},\label{6}
\end{eqnarray}

\begin{equation}
-8\pi {T^4}_4 = \frac{{\lambda}^{\prime}}{r} e^{-\lambda} + \frac{(1 - e^{-\lambda})}{r^2} = 8\pi \rho + \frac{q^2}{r^4},\label{7}
\end{equation}

where the prime denotes differential with respect to '$r$ and $E = \frac{1}{r^2}\,\int^r_0{4\,\pi\,r^2\,\sigma\,e^{\lambda/2}}=\frac{q}{r^2}$. However $\sigma$ is the charge density.   

\section{New Charged anisotropic solution for compact star:}

To solve the Einstein-Maxwell equations(\ref{5}-\ref{7}), we have considered following metric function $e^{\lambda(r)}$ and electric charge function $q(r)$ as:
\begin{equation}
e^{\lambda}=\frac{(1+2br^2+b^2r^4)}{(1+2ar^2+a^2r^4)} \label{8}
\end{equation}
where $a$ and $b$ are parameters with unit of $length^{-2}$. Recently Takisa et al.\cite{451} has obtained stellar model PSR J1614-2230 in the quadratic with quadratic equation of state by taking the metric function $e^{\lambda}=(1+ar^2)/(1+br^2)$. However the proposed metric function is also used by Newton et al.\cite{461} in class one. Also the gravitational metric function is singularity free at centre as $e^{\lambda(0)}=1$ and it is monotonically increasing with $r$. The behavior is shown by fig. (1).

\begin{figure}[h!]
    \centering
    \includegraphics[width=5.5cm]{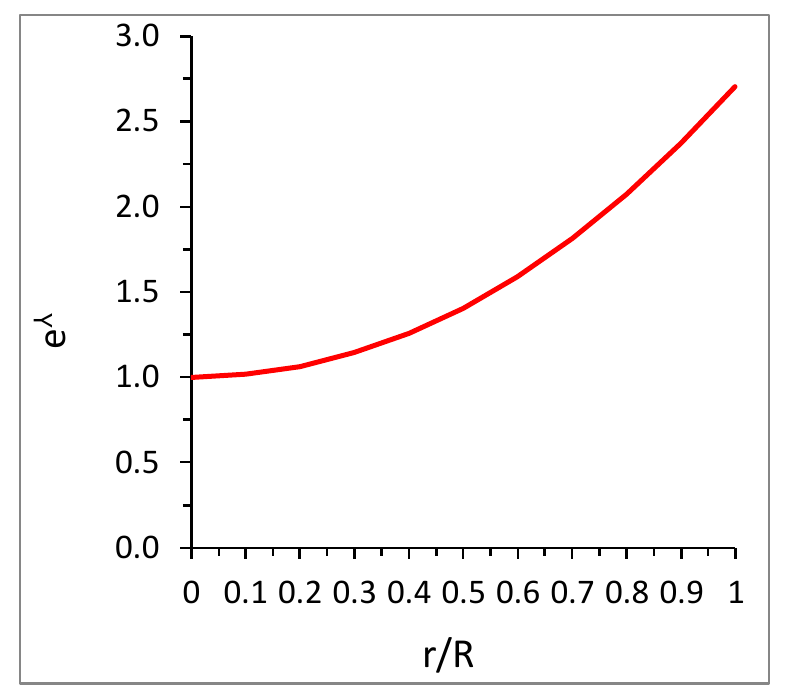}
\caption{Variation of metric function $e^{\lambda}$ with the fractional coordinate $r/R$ for $PSR 1937 +21$. For this graph, the numerical values of parameters are as follows: $a=0.002$, $b=0.0094$,  $R= 10.3142$, $M= 2.4115M_{\odot}$.}
\end{figure}

Now we have also considered the electric charged function to solve the system of equations which is given by

\begin{equation}
\frac{q^2}{r^4}=\frac{\alpha r^2}{(1+2br^2+b^2r^4)} \label{9}
\end{equation}

For physical validity of solution, The electric charge must vanish at centre and It should be increased with $r$. This is clear form fig.(2), the electric charge function is increasing away from centre and it is vanish at centre. The numerical values of $q$ have been shown in table 1. The amount of the charge on the boundary is $2.8605\times 10^{20}$ Coulomb (since the charge on boundary is $2.4535$, So for getting the charge in coulomb unit we have to multiply every value by the factor $1.1659\times 10^{20}$).

\begin{figure}
\centering
    \includegraphics[width=5.5cm]{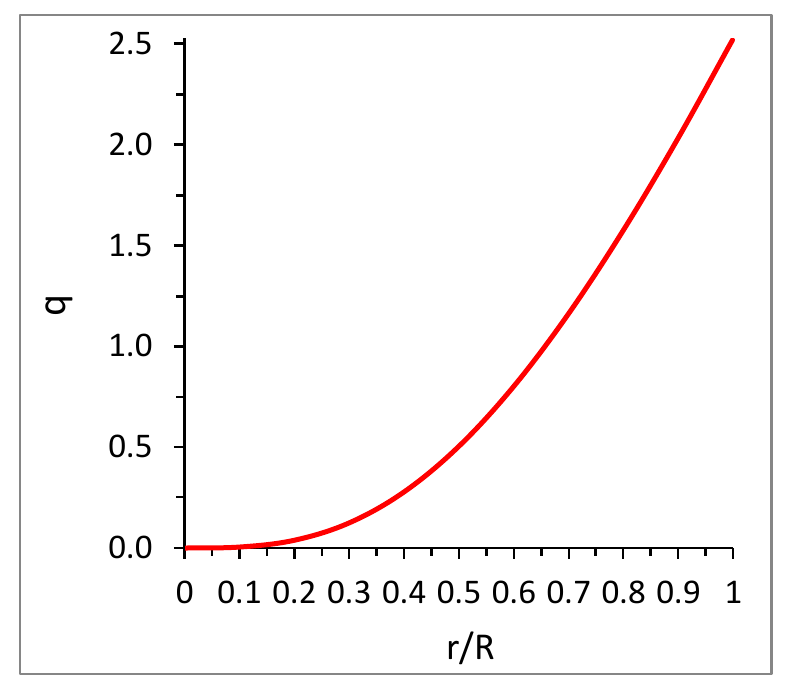}
\caption{Variation of electric charge $q$ with the fractional coordinate $r/R$ for $PSR 1937 +21$. For this graph, the values of parameters as: $b=0.0094$, $\alpha=0.00002$, $R= 10.3142$, $M= 2.4115M_{\odot}$.}
\end{figure}

By plugging the value of $\lambda$ and $q$ into eq.(7), we get the expression of energy density,
\begin{equation}
\rho=\frac{3b^2r^2 -\alpha r^2 +b^3r^4+2a(-3+br^2)-a^2r^2(5+br^2)+b(6-\alpha r^4)}{8 \pi \,(1+br^2)^3}
\label{10}
\end{equation}
The density profile has been shown in fig. (3), we observed the density is maximum at centre and it is decreasing throughout the star.

\begin{figure}
\centering
    \includegraphics[width=5.5cm]{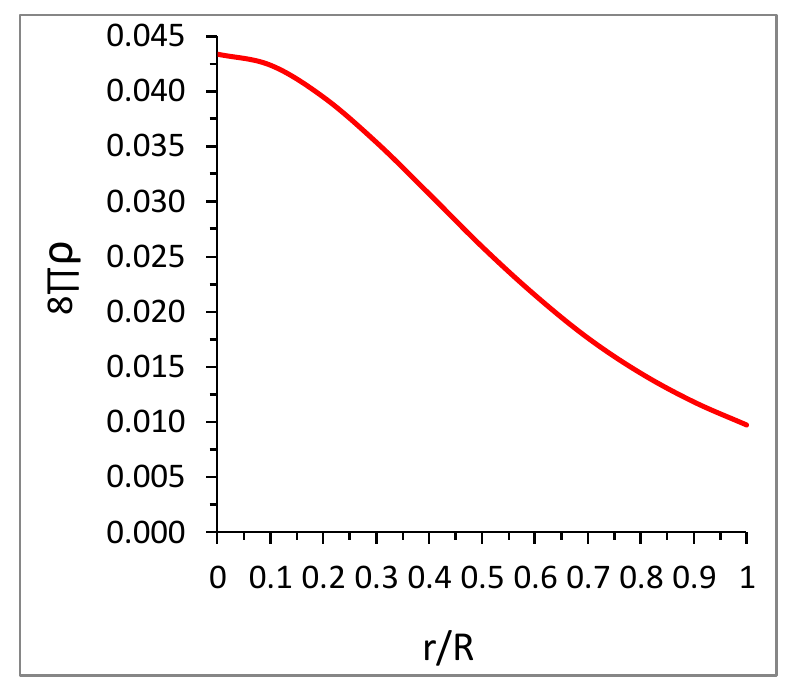}
\caption{Variation of density $\rho$ with the fractional coordinate $r/R$ for $PSR 1937 +21$. For this graph, the parameter values of constants are as follows: $a=0.002$, $b=0.0094$, $\alpha=0.00002$, $\beta=0.009$, $R= 10.3142$, $M= 2.4115M_{\odot}$.}
    \end{figure}

 Now if the electrically charged mass function is denoted as $m(r)$, then it can be written as:

 \begin{equation}
\frac{2\,m(r)}{r}=1-e^{-\lambda}+\frac{q^2}{r^2}  \label{11}
\end{equation}

By inserting the value of $\lambda(r)$ and $q(r)$ into eq.(11), we get:
\begin{equation}
m(r)=\frac{r^3}{2}\,\frac{[(-2a+2b-a^2 r^2 + b^2 r^2 + \alpha r^2)]}{(1 + b r^2)^2}  \label{12}
\end{equation}

\begin{figure}\centering
    \includegraphics[width=5.5cm]{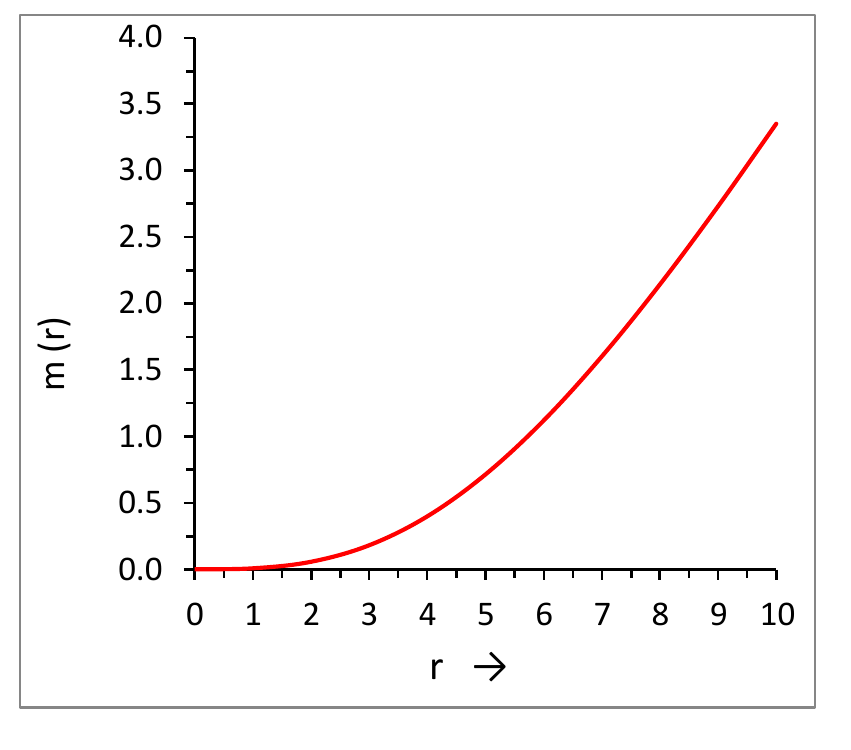}
\caption{Variation of mass function $m(r)$ with the radial vector  $r$. For this graph, the parameter values of constants are as follows: $a=0.002$, $b=0.0094$, $\alpha=0.00002$, $\beta=0.009$.}
\end{figure}

Since for any physical acceptable models the mass function must be increased with its radial coordinate and it should be zero at centre. As we can see from fig.(4) the mass function is monotonically increasing with $r/R$ and zero at centre.

Now our next aim to determine the metric function $\nu$ for examining other physical properties of the compact star. For this purpose we will consider expression for radial pressure which is positive finite inside the star and zero at boundary of star.  Let us consider expression of radial pressure $p_r$ as:

\begin{equation}
p_r=\frac{\beta\,(1-br^2)}{8\pi\,(1+2br^2+b^2r^4)} \label{13}
\end{equation}

\begin{figure}
\centering
    \includegraphics[width=5.5cm]{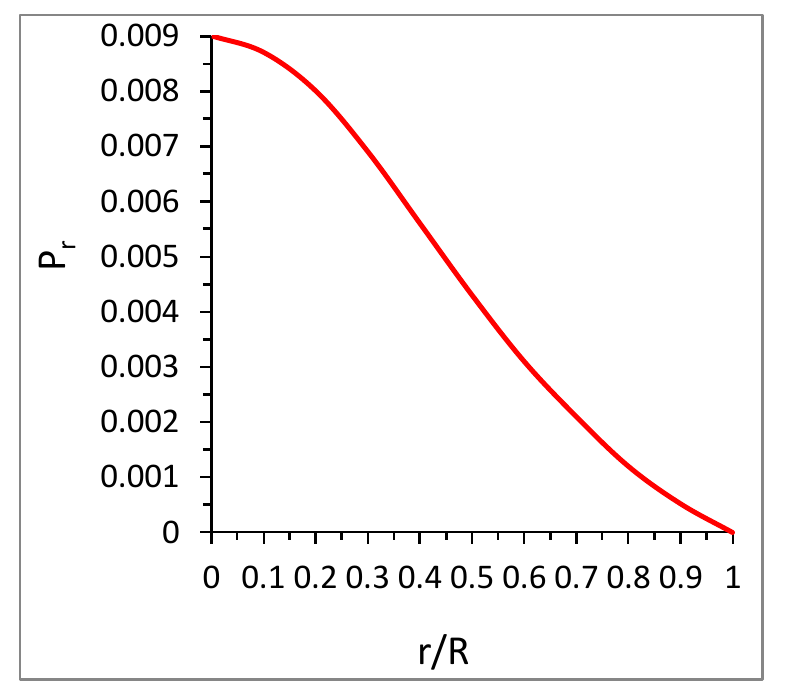}
\caption{Variation of radial pressure $P_r=8\pi p_r$ with the fractional coordinate $r/R$ for $PSR 1937 +21$. For this graph, the parameter values of constants are as follows: $a=0.002$, $b=0.0094$, $\alpha=0.00002$, $\beta=0.009$, $R= 10.3142$, $M= 2.4115M_{\odot}$.}
\end{figure}

 Since $p_r$ is zero at $r=\frac{1}{\sqrt{b}}$ i.e. $p_r\left(\frac{1}{\sqrt{b}}\right)=0$. Then $R=\frac{1}{\sqrt{b}}$ will be radius of the star. The behavior of radial pressure is shown in fig.(5), which shows that the radial pressure is maximum at centre and decreasing away from centre.

From Eq.(5) together with Eqs.(8), (9) and (13), we get,

\begin{equation}
\nu'=\frac{[(2b-2a+\beta)+(b^2-a^2-\beta\,b-\alpha)r^2]\,r}{\,(1+ar^2)^2} \label{14}
\end{equation}

We integrate above equation(\ref{14}) with respect to 'r', we get the metric function $\nu$ of the form as:

\begin{eqnarray}
\nu= \frac{(b^2-a^2-\beta\,b-\alpha)}{a^2}\,ln(1+ar^2)-\frac{[2ab-a^2-b^2+\beta(a+b)+\alpha]}{a^2(1+ar^2)}+ln A \label{15}
\end{eqnarray}
where, $A$ is constant of integration.

\begin{figure}
\centering
    \includegraphics[width=5.5cm]{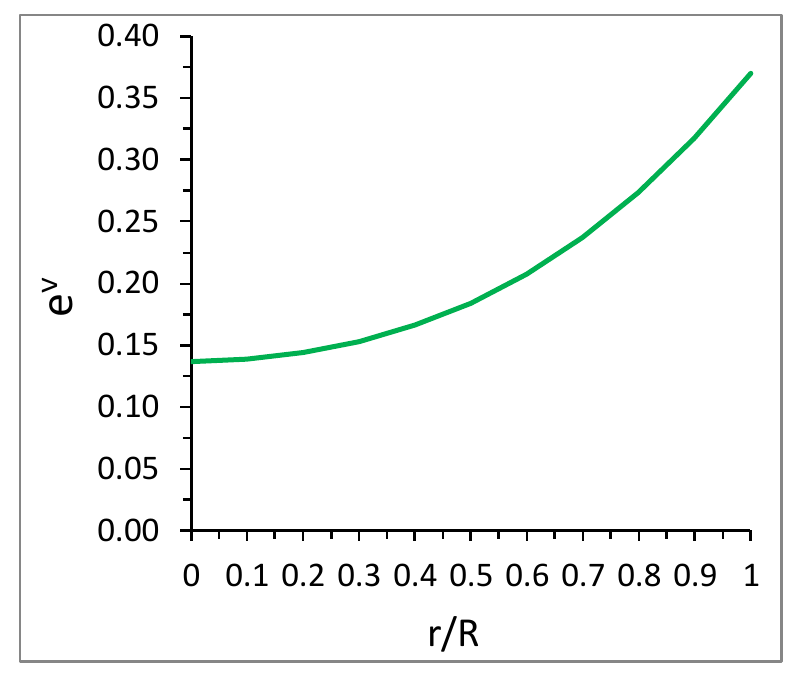}
\caption{Variation of metric function $e^{\nu}$ with the fractional coordinate $r/R$ for $PSR 1937 +21$. For this graph, the parameter values of constants are as follows: $a=0.002$, $b=0.0094$, $\alpha=0.00002$, $\beta=0.009$, $A=651.8666$, $R= 10.3142$, $M= 2.4115M_{\odot}$.}
\end{figure}

Let us discuss about the physical behavior of the metric function $\nu$. For any physically acceptable models, Lake\cite{2} has proved in general that the metric function $e^{\nu}$ must be increased with $r$ and it should be free from singularity at any point of the star. From the plot (6), the $e^{\nu}$ is monotonic increasing with increase of $r$ and it is free from singularity at every point.

Then the tangential pressure $p_t$ and anisotropic factor, $\Delta=p_t-p_r$, are determined as:
\begin{eqnarray}
p_t=\frac{p_{t1}+r^2[p_{t2}+p_{t3}+p_{t4}]}{ 32\pi\,(1+ar^2)^2 (1+br^2)^3} \label{16}
\end{eqnarray}

\begin{figure}
 \centering
    \includegraphics[width=5.5cm]{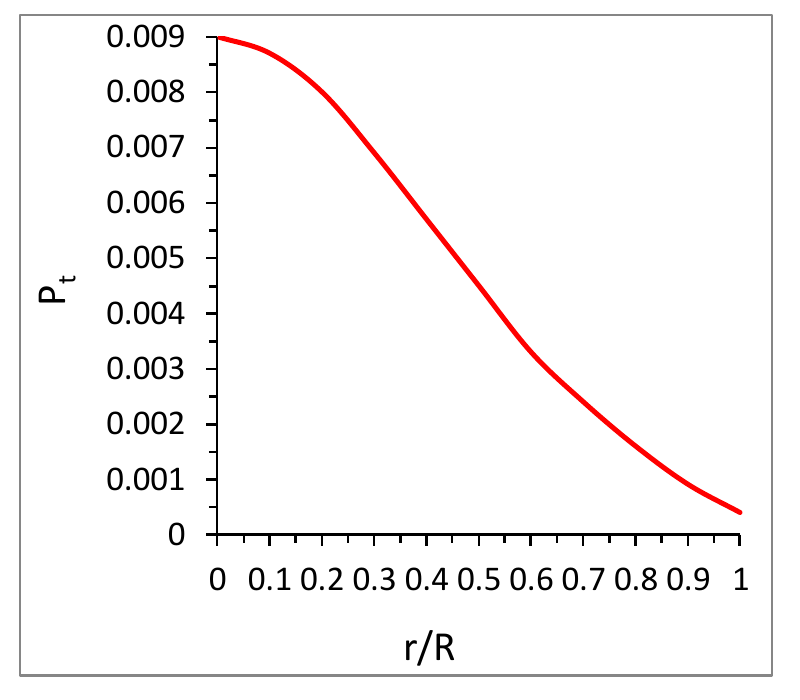}
\caption{Variation of tangential pressure $P_t=8\pi p_t$ with the fractional coordinate $r/R$ for $PSR 1937 +21$. For this graph, the parameter values of constants are as follows: $a=0.002$, $b=0.0094$, $\alpha=0.00002$, $\beta=0.009$, $R= 10.3142$, $M= 2.4115M_{\odot}$.}
\end{figure}

\begin{eqnarray}
 \Delta=\frac{\beta(-1+b^2r^4) (1+ar^2)^2 +p_{t1}+r^2[p_{t2}+p_{t3}+p_{t4}]}{32\pi  (1+ar^2)^2 (1+br^2)^3} \label{17}
\end{eqnarray}

$p_{t1}=\beta^2 r^2 (-1+b r^2)^2 (1 + b r^2)-2 \beta [-2 + a^2 r^4 + \alpha r^4 + 2 b^3 r^6 + b^4 r^8 + 2br^2(1+4ar^2+2a^2 r^4)+b^2r^4-b^2(a^2+\alpha)r^8]$;\\

$p_{t2}=16 a^3 r^2 + 5 b^4 r^4 + b^5 r^6 + a^4 r^4 (5 + b r^2) +\alpha (-12 + \alpha r^4) + b \alpha r^2 (-12 + \alpha r^4)$.\\

$p_{t3}=-2 b^3 r^2 (-6 + \alpha r^4) - 6 b^2 (-2 + \alpha r^4) - 8 a [b^2 r^2 + 2 \alpha r^2 + b (3 + \alpha r^4)]$,\\

$p_{t4}=-2 a^2 [-6 + 5 b^2 r^4 + 3 \alpha r^4 + b^3 r^6 + b r^2 (10 + \alpha r^4)]$,\\

 Since plot for $p_t$ is shown in fig.(7). it is clear from the figure that $p_t$ is monotonically decreasing throughout the star. While the anisotropy factor is zero at centre and monotonic increasing away from centre. This features of anisotropy allows to construct more realistic compact objects. The behavior of anisotropy factor can be seen by fig. (8).

\begin{figure}
\centering
    \includegraphics[width=5.5cm]{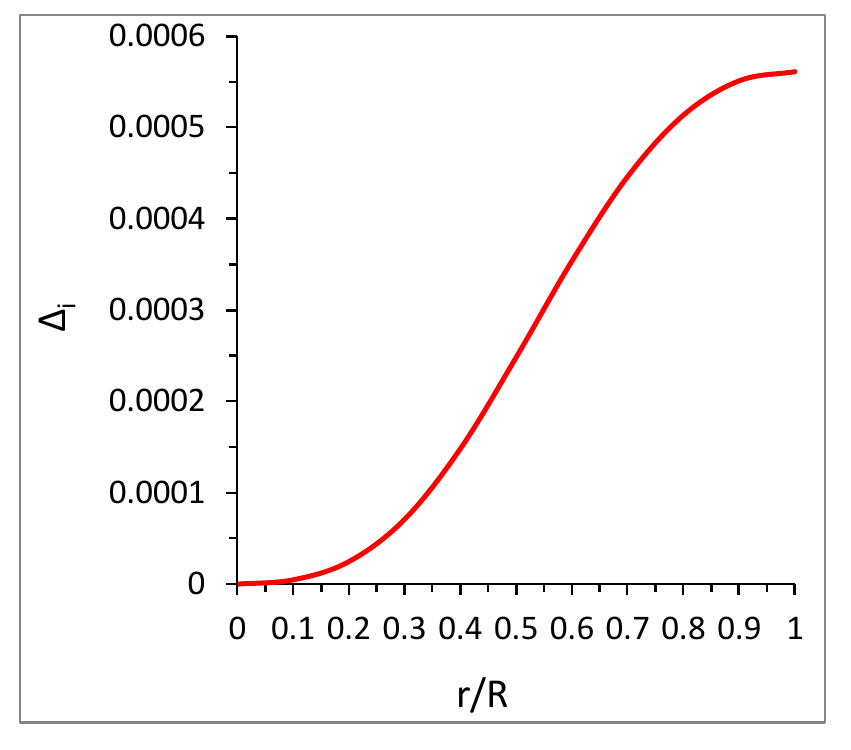}
\caption{Variation of anisotropic factor $\Delta$ with the fractional coordinate $r/R$ for $PSR 1937 +21$. For this graph, the parameter values of constants are as follows: $a=0.002$, $b=0.0094$, $\alpha=0.00002$, $\beta=0.009$, $R= 10.3142$, $M= 2.4115M_{\odot}$.}
\end{figure}

\section{Equation of state for present charged compact star:}
The equation of state is an important relation in pressure and density for any realistic matter and it is necessary to provide the equation of state (EOS) $\rho=f(p_i)$ to model of the present charged compact star:

From Eqs.(10) and (13), we get:

\begin{equation}
\rho=\frac{[\,3b^2\,f_1 -\alpha\,f_1 +b^3\,{f_1}^2+f_2\,]\,p_r}{\beta\,(1-f_b)\,(1+f_b)} \label{18}
\end{equation}

\begin{eqnarray}
\rho=\frac{4\,(1+\,f_{a})^2\,[\,3b^2\,f_1 -\alpha\,f_1 +b^3\,{f_1}^2+f_2\,]\,p_t}{\beta^2 f_1 (-1+f_b)^2 (1 + f_b)+f_{3}+f_{1}[f_{4}+f_{f5}+f_{6}]} \label{18a}
\end{eqnarray}

where,

$f_1=\frac{-(16\pi\,p_r+\beta)+\sqrt{64\,\pi\,p_r\,\beta+\beta^2}}{16\pi\,b\,p_r}$,\\

$f_2=2a(-3+b\,f_1)-a^2\,f_1(5+b\,f_1)+b(6-\alpha {f_1}^2)$.\\

$f_a=a\,f_1$, $f_b=b\,f_1$,\\

$f_3=-2 \beta [-2 + {f_a}^2 + \alpha {f_1}^2 + 2 {f_b}^3 + {f_b}^3 + 2f_b\,(1+4f_a+2{f_b}^2)+{f_b}^2-b^2(a^2+\alpha){f_1}^4]$;\\

$f_4=16 a^2 f_a + 5 b^2 {f_b}^2 + b^2 {f_b}^3 + a^2 {f_a}^2 (5 + {f_b}) +\alpha (-12 + \alpha {f_1}^2) + b \alpha f_1 (-12 + \alpha {f_1}^2)$.\\

$f_5=-2 b^2 f_b (-6 + \alpha {f_1}^2) - 6 b^2 (-2 + \alpha {f_1}^2) - 8 a [b f_b + 2 \alpha f_1 + b (3 + \alpha {f_1}^2)]$,\\

$f_6=-2 a^2 [-6 + 5 {f_b}^2 + 3 \alpha {f_1}^2 + {f_b}^3 + f_b\, (10 + \alpha {f_1}^2)]$.\\

The above relation (Eq.\ref{18}) implies that the density is purely function of radial and tangential pressure both. So it may represent an equation of state (EOS) for present charged compact star.

\section{Boundary condition for determining the constants:}
To determine arbitrary constants of the anisotropic charged fluid solution we must join the interior of metric (4) to the Reissner-Nordstr{\"o}m metric at boundary of the star ($r=R$). The Reissner-Nordstr{\"o}m metric is given as:
\begin{eqnarray}
ds^{2} =-\left( 1-\frac{2M}{r} +\frac{Q^{2} }{r^{2} } \right)^{-1} dr^{2} +\left( 1-\frac{2M}{r} +\frac{Q^{2} }{r^{2} } \right) dt^{2}
-r^{2} (d\theta ^{2} +\sin ^{2} \theta d\phi ^{2})  .   \label{19}
\end{eqnarray}
The joining of metric condition demands the continuity of $e^{\lambda }$, $ e^{\nu } $ and $Q$ across the boundary $r=R$, so that

\begin{equation}
e^{-\lambda (R)} =1-\frac{2M}{R} +\frac{Q^{2} }{R^{2} },
\label{20}
\end{equation}

\begin{equation}
e^{\nu (R)} = 1-\frac{2M}{R} +\frac{Q^{2} }{R^{2} }, \label{21}
\end{equation}

\begin{equation}
q(R)=Q, \label{22}
\end{equation}
However It is also requires that the radial pressure, $p_r$ , should be zero at boundary (continuity of second fundamental form at boundary), so that:
 \begin{equation}
R=\sqrt{\frac{1}{b}},
\end{equation}
The conditions (20) and (21), give the value of $A$ as:
\begin{equation}
A=\frac{(1+aR^2)^{\frac{5a^2-b^2+\beta\,b+\alpha)}{2a^2}}}{(1+bR^2)^2}\,\, e^{\frac{[\,2ab-a^2-b^2+\beta(a+b)+\alpha\,]}{2a^2(1+aR^2)}}
\end{equation}

\section{Physical features of the charged anisotropic models:}

\subsection{Causality condition:}
For any physical acceptable anisotropic charged fluid models, The speed of sound must be less than the speed of light inside stars. Then it requires to satisfy the following conditions $V_r=\sqrt{dp_r/d\rho}<1$ and $V_t=\sqrt{dp_t/d\rho}<1$. From fig.(9), we have observed that the radial and transverse speed of sound are always less than the speed of light.

\begin{figure}\centering
    \includegraphics[width=5.5cm]{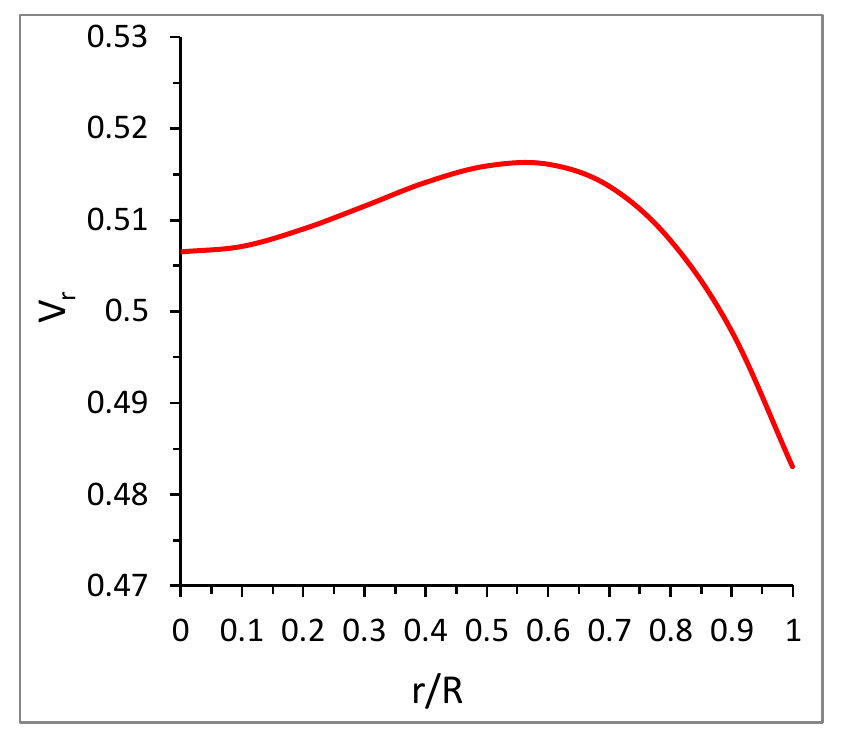} \includegraphics[width=5.5cm]{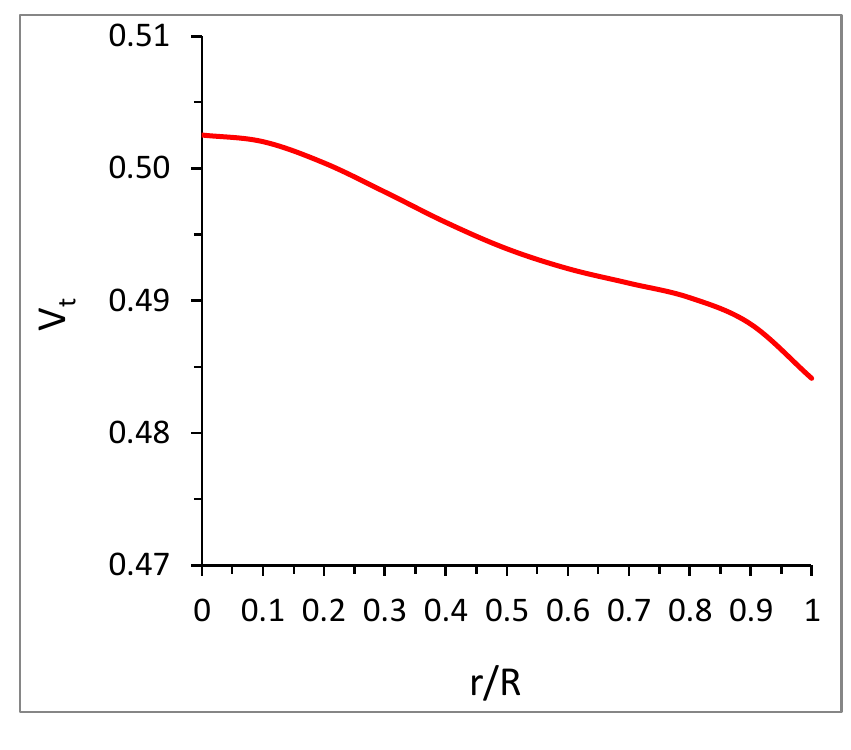}
\caption{Variation of radial velocity (left panel) and tangential velocity (right panel) with the fractional coordinate $r/R$ for $PSR 1937 +21$. For this graph, the parameter values of constants are as follows: $a=0.002$, $b=0.0094$, $\alpha=0.00002$, $\beta=0.009$, $R= 10.3142$, $M= 2.4115M_{\odot}$.}
\end{figure}

\subsection{Stability condition:}
\subsubsection{Adiabatic index:}
Heintzmann and Hillebrandt\cite{47} have proposed that neutron
star models are stable with anisotropic equation of state if $\Gamma_i >
4/3$.  Also in the Newton's theory of gravitation, it is well demonstrate that there
has no upper mass limit when the equation of state has an adiabatic index $\Gamma_i > 4/3$.
The adiabatic index corresponding radial and transverse pressure are defined as:

\begin{equation}
\Gamma_r = \left(\frac{\rho+p_r}{p_r}\right) \left(\frac{dp_r}{d\rho}\right),\label{25}
\end{equation}
\begin{equation}
\Gamma_t = \left(\frac{\rho+p_t}{p_t}\right) \left(\frac{dp_t}{d\rho}\right),\label{26}
\end{equation}

\begin{figure}\centering
    \includegraphics[width=5.5cm]{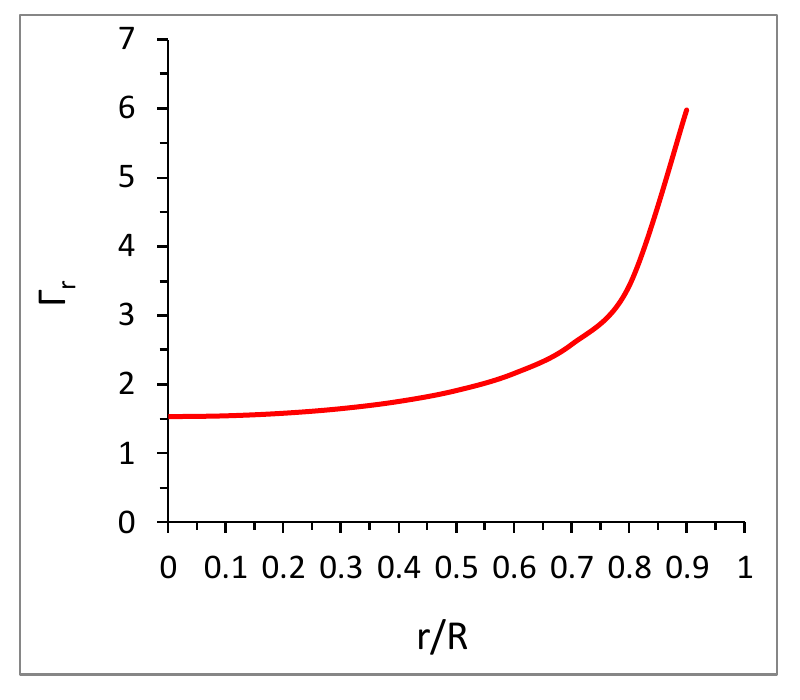} \includegraphics[width=5.5cm]{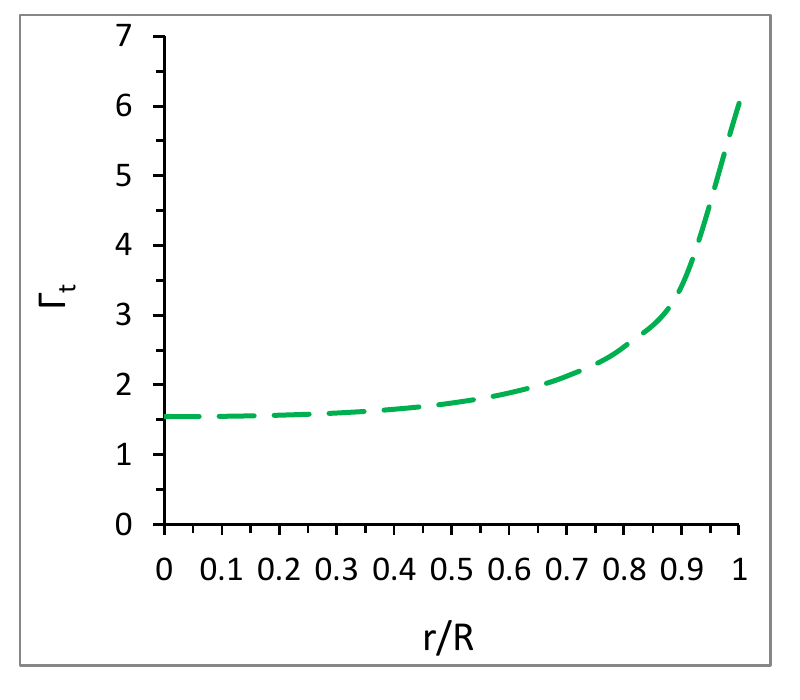}
\caption{Variation of adiabatic index corresponding to radial pressure (left panel) and tangential pressure (right panel) with the fractional coordinate $r/R$ for $PSR 1937 +21$. For this graph, the parameter values of constants are as follows: $a=0.002$, $b=0.0094$, $\alpha=0.00002$, $\beta=0.009$, $R= 10.3142$, $M= 2.4115M_{\odot}$.}
\end{figure}

As we can see from fig.(10) the radial and tangential adiabatic index have values more than $4/3$ everywhere inside the star. Then our charged fluid models are stable.

\subsubsection{Herrera Cracking concept:}
In the charged anisotropic compact star models, to examine the stability of the model we plot the radial
($V_{r}^2=dp_r/d\rho$) and transverse ($V^2_{t}=dp_t/d\rho$) sound speeds in Fig.11. It shows that $V^2_r$ and $V^2_t$ satisfy the
inequalities $0\leq V_{r}^2 \leq 1$ and $0\leq V_{t}^2 \leq 1$ everywhere within the charged anisotropic stellar object \cite{48,49}.

\begin{figure} \centering
    \includegraphics[width=5cm]{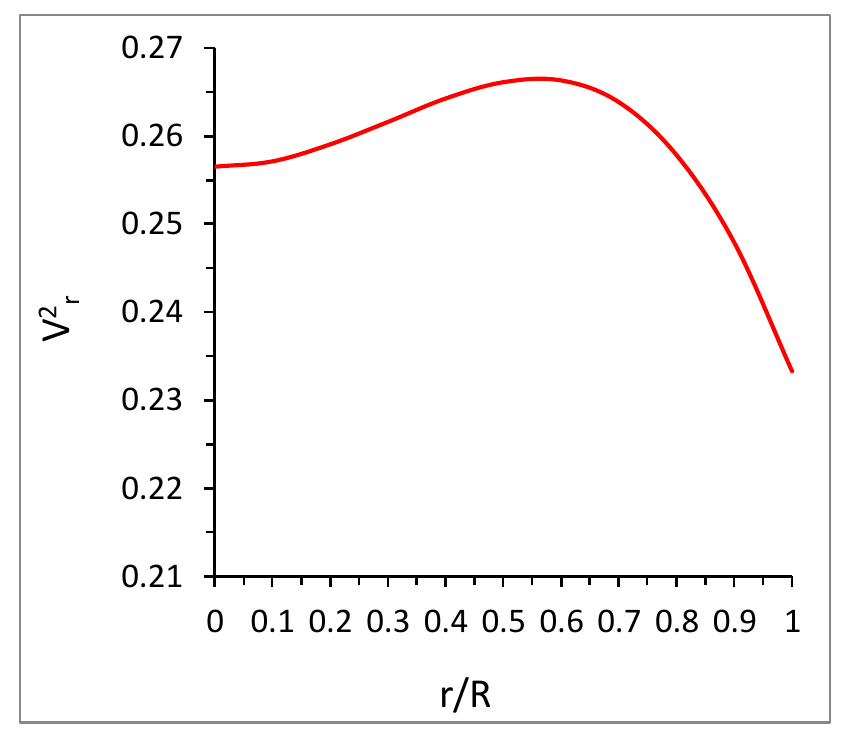} \includegraphics[width=5cm]{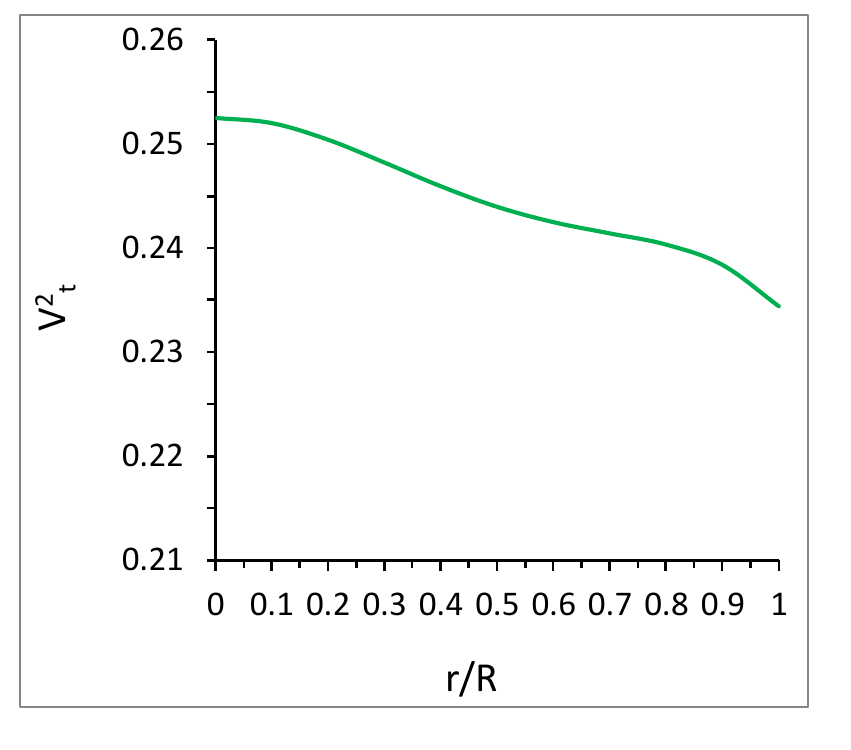}
\caption{Variation of square of radial velocity $V^2_r$ and tangential velocity $V^2_t$ with the fractional coordinate $r/R$ for $PSR 1937 +21$. For this graph, the parameter values of constants are as follows: $a=0.002$, $b=0.0094$, $\alpha=0.00002$, $\beta=0.009$, $R= 10.3142$, $M= 2.4115M_{\odot}$.}
\end{figure}

Also to discuss about the stability of anisotropic charged compact star whether the anisotropic charged matter distribution is stable or not, we will use the concept of Herrera\cite{48} cracking concept which states that the region is potentially stable where the radial velocity of sound is greater than the tangential velocity of sound.

\begin{figure}\centering
    \includegraphics[width=5.5cm]{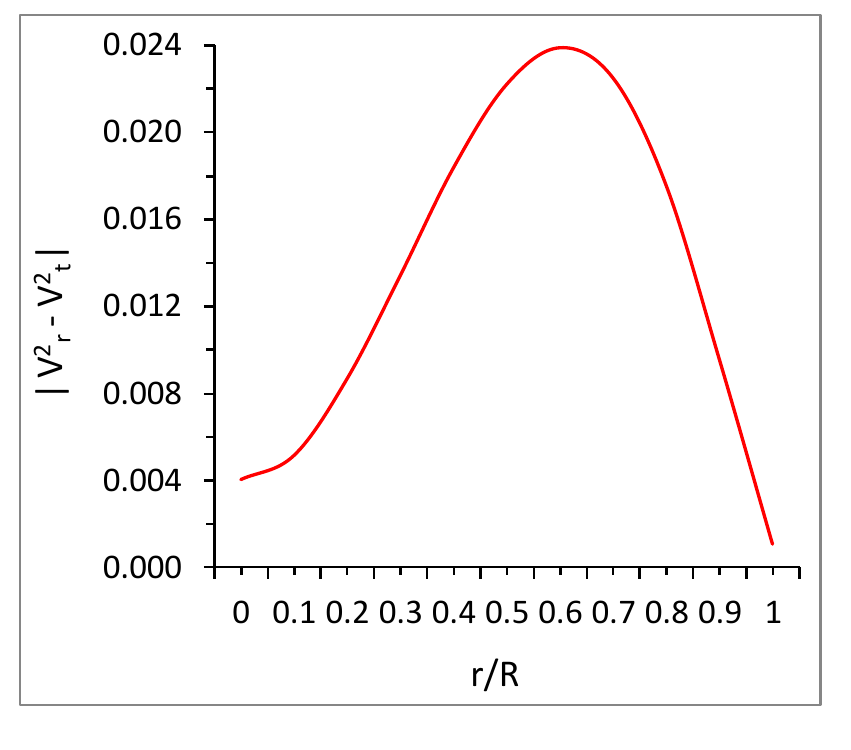}
\caption{Variation of $|V^2_r-V^2_t|$ with the fractional coordinate $r/R$ for $PSR 1937 +21$. For this graph, the parameter values of constants are as follows: $a=0.002$, $b=0.0094$, $\alpha=0.00002$, $\beta=0.009$, $R= 10.3142$, $M= 2.4115M_{\odot}$.}
\end{figure}

\subsubsection{Tolman-Oppenheimer-Volkoff (TOV) equation:}

Let us write the generalized Tolman-Oppenheimer-Volkoff (TOV) equation to examine the stability of the model under stable equilibrium configuration as:
\begin{equation}
-\frac{M_G(\rho+p_r)}{r^2}e^{(\lambda-\nu)/2}-\frac{dp_r}{dr}+
\sigma \frac{q}{r^2}e^{\lambda/2} + \frac{2}{r}(p_t-p_r)=0,
\end{equation}
where $M_G$ is the gravitational mass within the radius $r$ and $\sigma$ is the charged density, which are defined as:
given by
\begin{equation}
M_G(r)=\frac{1}{2}r^2 \nu^{\prime}e^{(\nu - \lambda)/2}\,\, and \,\, \sigma=\frac{e^{-\lambda/2}}{4\pi} \frac{dq}{dr}
\end{equation}

On inserting the value of $M_G$ and $\sigma$ into above equation, we get

\begin{equation}
-\frac{1}{2} \nu^{\prime}(\rho+p_r)-\frac{dp_r}{dr}+ \sigma
\frac{q}{4\pi\,r^2} \frac{dq}{dr}+\frac{2}{r}(p_t-p_r)=0.
\end{equation}

The above TOV equation describes the equilibrium condition for a
charged anisotropic fluid subject to gravitational force ($F_g$),
hydrostatic force ($F_h$), electric charge force ($F_e$) and anisotropic stress
($F_a$) so that:
\begin{equation}
F_g+F_h+F_e+F_a=0,
\end{equation}

where the explicit form of above these forces can be defined as:
\begin{equation}
F_g=-\frac{1}{2} \nu^{\prime}(\rho+p_r)=-\frac{F_{g1}}{32\pi\,(1+ar^2)^2}\left[\frac{F_{g2}}{(1+br^2)^3}\right],
\end{equation}

\begin{equation}
F_h=-\frac{dp_r}{dr},
\end{equation}

\begin{equation}
F_e=\frac{q}{4\pi\,r^2} \frac{dq}{dr}=\frac{\alpha\,r^3}{4\,\pi}\frac{(3+br^2)}{(1+br^2)^3},
\end{equation}

\begin{equation}
F_a=\frac{2}{r}(p_t-p_r)
\end{equation}

where,

$F_{g1}=\frac{r\,[2b-2a+\beta)+(b^2-a^2-\beta\,b-\alpha)r^2]}{2(1+ar^2)^2}$

$F_{g2}=\beta+6b+3b^2r^2-\alpha r^2-\beta b^2 r^4+b^3 r^4-b \alpha r^4 +2a(-3+br^2)-a^2 r^2 (5+br^2)$

We have shown the figure Fig.13 for the TOV equations. From
this plot it is observed that the system is in static equilibrium
under four different forces, e.g. gravitational, hydrostatic,
electric and anisotropic forces. However, strong gravitational force is counter balanced by
the joint action of hydrostatic and electric forces and anisotropic force. Also as we can see that the effect of anisotropic stress has negligible effect to  balance this mechanism.

\begin{figure}\centering
    \includegraphics[width=5.5cm]{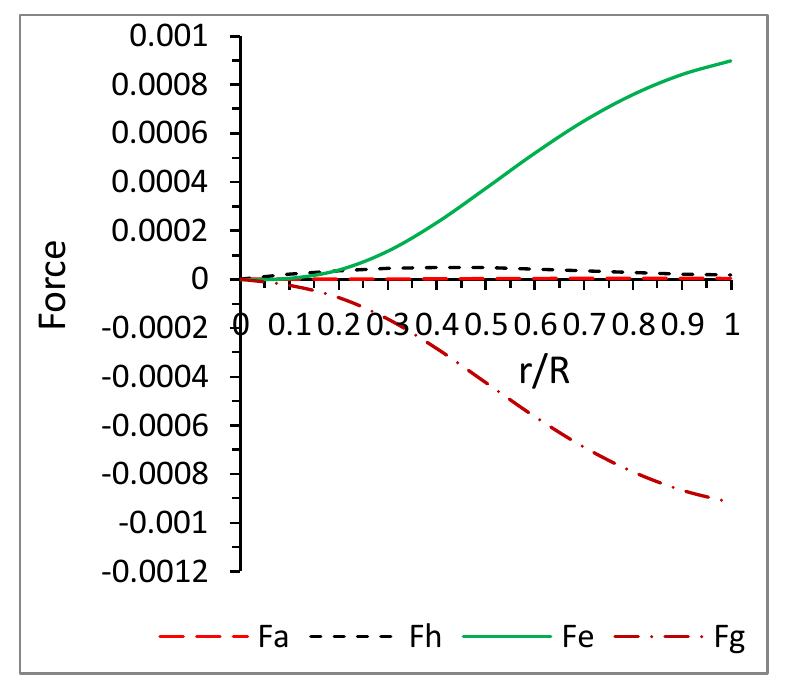}
\caption{Variation of different forces with the fractional coordinate $r/R$ for $PSR 1937 +21$. For this graph, the parameter values of constants are as follows: $a=0.002$, $b=0.0094$, $\alpha=0.00002$, $\beta=0.009$, $R= 10.3142$, $M= 2.4115M_{\odot}$.}
\end{figure}

\subsection{Maximum limit of mass-radius ratio:}
The maximum limit of mass-radius ratio is proposed by Buchdahl\cite{4} for isotropic compact star which satisfy the inequality $2M/R \leq 8/9$. This inequality imply that mass of static isotropic compact star cannot be arbitrary large. Also Bohmer and Harko\cite{50} provided the lower limit of mass-radius ratio for the charged compact star model with $ Q < M$,

\begin{equation}
\frac{Q^2}{R^2}
\left(\frac{18R^2+{Q^2}}{12R^2+{Q^2}}\right) \leq
\frac{2M}{R},\label{35}
\end{equation}

In addition to that the upper bound of the mass of charged sphere was generalized by
Andréasson\cite{51} and proved that
\begin{equation}
\sqrt{M} \leq \frac{\sqrt{R}}{3} + \sqrt{\frac{R^2+3Q^2}{9R}}.\label{36}
\end{equation}

Then the range of mass $M$ can be given by the following inequality:
\begin{equation}
\frac{Q^2}{2R}
\left(\frac{18R^2+{Q^2}}{12R^2+{Q^2}}\right) \leq
M \leq \left[\frac{\sqrt{R}}{3} + \sqrt{\frac{R^2+3Q^2}{9R}}\right]^2
\end{equation}

\subsection{Surface red-shift}

Let us write the effective mass for the charged compact star as:
\begin{equation}
M_{eff}=4\pi\int_{0}^{R}\left(\rho +\frac{E^2}{8\pi}\right) r^2
dr= \frac{1}{2}R[1-e^{-\lambda(R)}], \label{38}
\end{equation}
Then the compactness factor $u$ for above mass is defined as:
\begin{equation}
u=\frac{M_{eff}}{R}
\end{equation}

Now the surface red-shift for above compactness $u$ can be written as:

\begin{equation}
Z_s(R) = \sqrt{\frac{1}{[1-2u]}} - 1 = e^{\lambda(R)/2} - 1. \label{40}
\end{equation}

As we can see from the Eq.(40) the surface redshift function dependent on compactness factor $u$. Since $Z_r(R)$ is monotonically increasing with increase of  $u$. Then surface redshift function cannot have any arbitrary values as $u$ is always less than $4/9$ (Buchdahl limit). The variation of redshift of the charged anisotropic solution is shown by fig.14. We observed from this figure the redshift is maximum at centre and minimum at boundary of star.

\begin{figure} \centering
    \includegraphics[width=5.5cm]{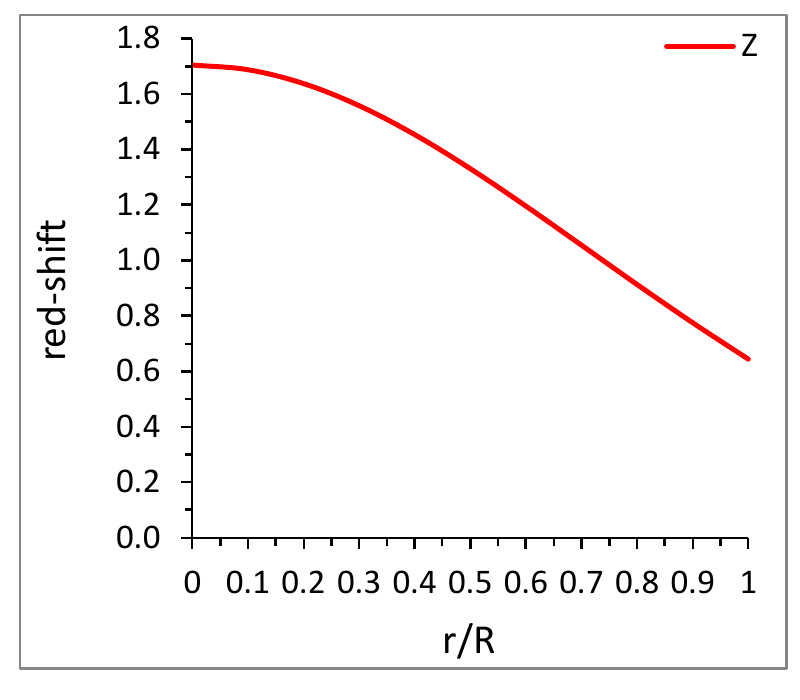}
\caption{Variation of red-shift $Z$ with the fractional coordinate $r/R$ for $PSR 1937 +21$. For this graph, the parameter values of constants are as follows: $a=0.002$, $b=0.0094$, $\alpha=0.00002$, $\beta=0.009$, $R= 10.3142$, $M= 2.4115M_{\odot}$.}
\end{figure}

\begin{table}
\caption{ Values of different physical parameters of PSR 1937 +21 for $a$=0.002, $b$ = 0.0094, $\alpha$ =0.00002, $\beta$ =0.009, $M = 2.4115M_{\odot}$, R = 10.3142 Km, $P_r=8\pi p_r$, $P_t=8\pi p_t$, $D=8\pi \rho$, $\Delta_i=8\pi\Delta$}
	\label{Table1}
	\begin{tabular}{@{}lrrrrrrrrrr@{}} \hline
		$r/R$ & $P_r$ & $P_t$  & $D$    & $q$      & $\Delta_i$  & $V_r$  & $V_t$  &$e^{\nu}$ & $e^{\lambda}$& $Z$ \\\hline
		0.0 & 0.009 & 0.009 & 0.0444    & 0.0000   & 0.0000               & 0.5065 & 0.5025 & 0.1353 & 1      & 1.7185 \\ \hline
		0.1 & 0.0087 & 0.0087 & 0.0434  & 0.0049   & 4.74$\times 10^{-6}$ & 0.5071 & 0.502 & 0.137 & 1.0158& 1.7014  \\ \hline
		0.2 & 0.008 & 0.008 & 0.0405    & 0.0377   & 2.47$\times 10^{-5}$ & 0.5090 & 0.5004 & 0.1423 & 1.0634 & 1.6512  \\ \hline
		0.3 & 0.0069 & 0.0069 & 0.0363  & 0.1216  & 7.11$\times 10^{-5}$ & 0.5115 & 0.4982 & 0.1513 & 1.1439 & 1.5713  \\ \hline
		0.4 & 0.0056 & 0.0057 & 0.0314  & 0.2707   & 1.48$\times 10^{-4}$ & 0.5141 & 0.4959 & 0.1644 & 1.2585 & 1.4666  \\ \hline
		0.5 & 0.0043 & 0.0045 & 0.0265  & 0.4907   & 2.48$\times 10^{-4}$ & 0.5159 & 0.4939 & 0.1821 & 1.4087 & 1.3432  \\ \hline
		0.6 & 0.0031 & 0.0033 & 0.022   & 0.7794   & 3.54$\times 10^{-4}$ & 0.5161 & 0.4924 & 0.2052 & 1.5958 & 1.2073  \\ \hline
		0.7 & 0.0021& 0.0024 & 0.0181   & 1.1296   & 4.46$\times 10^{-4}$ & 0.5137 & 0.4913 & 0.2345 & 1.8207 & 1.0652  \\ \hline
		0.8 &0.0012 & 0.0016 & 0.0148   & 1.532   & 5.13 $\times 10^{-4}$ & 0.5078 & 0.4902 & 0.2707 & 2.0835 & 0.9221  \\ \hline
        0.9 & 0.00052 & 0.0009 &0.121   & 1.9764  & 5.51$\times 10^{-4}$  &  0.4979 & 0.4882 & 0.3148 & 2.3837 & 0.7823  \\ \hline
        1.0 &0.000& 0.0004 & 0.0097    & 2.4535  & 5.61$\times 10^{-4}$   &  0.483 & 0.4841  & 0.3677 & 2.7196 & 0.6491  \\ \hline
	\end{tabular}
\end{table}

\begin{table}
\caption{ Central density, surface and central pressure for PSR 1937 +21 for the above parameter values of Tables 1}
	\label{Table2}
	\begin{tabular}{@{}lrrrrrr@{}} \hline
		Compact star & Central Density & Surface density & Central pressure & Mass & Radius(Km)& 2M/R \\
		candidates & $gm/cm^{3} $ & $gm/cm^{3}$ & $dyne/cm^{2}$ & $(M_{\odot})$ & \\ \hline
		PSR 1937 +21  & 2.3843$\times 10^{15} $ & 5.3157$\times 10^{14} $ & 4.3506$\times 10^{35}$ & 2.4115 & 10.3142 & 0.6888 \\ \hline
	\end{tabular}
\end{table}
\section{Conclusion:}

In the present article, we have obtained the charged anisotropic compact star models via radial pressure. The Einstein- Maxwell field equations are solved by using the metric function $\lambda$ (which is generalized form of earlier work by Maurya et al.\cite{52}), radial pressure and electric charge function. The obtained generic function $\nu$ is physically valid due to its monotonic increasing nature away from the centre and $\nu(0) \ne 0$. The charged anisotropic solution having the following features:\\
(i). The physical requirements of mass function has been investigated. Recently Lake\cite{2} and Maurya et al.\cite{35} have proved in general the mass function should zero at centre and must have increasing nature throughout from the centre. In present article, the investigated mass function is satisfying above physical requirements (Fig. 4).\\
(ii). The energy density of the models is positive and decreasing away from the centre, which can be seen from Fig.(3).\\
(iii). The radial pressure at centre is $p_r(0)=\beta/8 \pi$, which is positive. However it is zero at boundary i.e. $p_r(R)=p_r(\frac{1}{\sqrt{b}})=0$, which gives radius of the star  $R=\frac{1}{\sqrt(b)}$. From the figure 5, it is observed that $p_r$ is decreasing outward and zero on the boundary.\\
(iv). The metric function $e^{\nu}$ and $e^{\lambda}$ at centre are: $e^{\lambda(0)}=1$ and $e^{\nu(0)}=A\,e^{\frac{[a^2+b^2-2ab-\beta(a+b)-\alpha]}{2a^2}}$. As we can observe that both metric functions are positive and non-singular at centre and also both have increasing nature (Figs. 1 and 6).\\
(v). The anisotropic factor $\Delta=p_t-p_r$ always positive inside the star. This implies that $p_t > p_r$ and the force $F_a=\frac{2(p_t-p_r)}{r}$ is directed outward, which allows to the construction of more realistic compact objects \cite{16}. The behavior is shown in Fig. 8. \\
(vi). It is observed from fig.(9) that the speed of sound is less than speed of light throughout inside the star. However the radial velocity has no monotonic decreasing nature while tangential velocity has monotonic decreasing nature inside the star.\\
(vii). Stability conditions: we have investigated the stability of the charged compact star model via three different ways like Herrera cracking concept, adiabatic index and TOV equations. According to the Herrera cracking concept, the region is potentially stable where the radial speed of sound is greater than the tangential velocity and there is no change in sign of $V^2_r-V^2_t$ everywhere inside the star, which can be observed from the fig. 11 and fig. 12. Moreover the sufficient condition of stability for any models has an adiabatic index $\Gamma_i > 4/3$. It is observed from Fig.(10), $\Gamma_r $ and $\Gamma_t $ have value more than $4/3$ everywhere inside the star. Now let us examine the equilibrium condition for present models, we observe from Fig.(13) that the gravitational force $F_g$ is counterbalance by joint action of hydrostatic force $F_h$ and electric force $F_e$, while the anisotropic stress has negligible effect to balance this system. This implies that our system is in equilibrium stage.\\
(viii). The surface red-shift of the model depends upon the compactness factor $u$. if the value of compactness will increase then corresponding surface red-shift will also increase. This imply that redshift of the models can not be arbitrary large. The surface redshift of the model is turns out to be $Z_s=0.6491$ which is good in agreement for present charged compact star models \cite{50,53,54}.
As the final comment, the author has presented charged anisotropic solution which is representing the known compact star with their observed mass and radius (Table 2). We have also obtained the equation of state (EOS) for the present compact star model, which is most important physical property to describe structure of any realistic matter. As we can see from Eqs.(24) and (25) the density is purely function of pressers. Hence we conclude that this approach may help to describe the structure of compact star. Thus our present charged anisotropic solution might have some astrophysical application in future.

\end{document}